\documentclass[3p,times]{elsarticle}

\makeatletter
\def\ps@pprintTitle{%
 \let\@oddhead\@empty
 \let\@evenhead\@empty
 \let\@oddfoot\@empty
 \let\@evenfoot\@empty}
\makeatother

\usepackage{amssymb}
\usepackage[figuresright]{rotating}

\usepackage{hyperref}
\usepackage{subcaption} % for subfigure environment
\usepackage{booktabs}   % for \toprule, \midrule, \bottomrule
\usepackage{multirow}   % for \multirow in the table

\begin{document}

\begin{frontmatter}

%% Title, authors and addresses

%% use the tnoteref command within \title for footnotes;
%% use the tnotetext command for the associated footnote;
%% use the fnref command within \author or \address for footnotes;
%% use the fntext command for the associated footnote;
%% use the corref command within \author for corresponding author footnotes;
%% use the cortext command for the associated footnote;
%% use the ead command for the email address,
%% and the form \ead[url] for the home page:
%%
%% \title{Title\tnoteref{label1}}
%% \tnotetext[label1]{}
%% \author{Name\corref{cor1}\fnref{label2}}
%% \ead{email address}
%% \ead[url]{home page}
%% \fntext[label2]{}
%% \cortext[cor1]{}
%% \address{Address\fnref{label3}}
%% \fntext[label3]{}

% \dochead{}
%% Use \dochead if there is an article header, e.g. \dochead{Short communication}
%% \dochead can also be used to include a conference title, if directed by the editors
%% e.g. \dochead{17th International Conference on Dynamical Processes in Excited States of Solids}

\title{SkinGenBench: Generative Model and Preprocessing Effects for Synthetic Dermoscopic Augmentation in Melanoma Diagnosis}

%% use optional labels to link authors explicitly to addresses:
%% \author[label1,label2]{<author name>}
%% \address[label1]{<address>}
%% \address[label2]{<address>}

\author[label1]{N. A. Adarsh Pritam}
\ead{padarshDS24@arts.alliance.edu.in}

\author[label1]{Jeba Shiney O}
\ead{jeba.shiney@alliance.edu.in}

\author[label2,label3]{Sanyam Jain\corref{cor1}}
\ead{au775886@uni.au.dk}

\cortext[cor1]{Corresponding author}

\address[label1]{Alliance School of Advanced Computing, Alliance University, India}
\address[label2]{Department of Computer Science and Communication, Østfold University College, 1783 Halden, Norway}
\address[label3]{Department of Dentistry and Oral Health, Aarhus University, 8000 Aarhus, Denmark}

\begin{abstract}
This work introduces SkinGenBench, a systematic biomedical imaging benchmark that investigates how preprocessing complexity interacts with generative model choice for synthetic dermoscopic image augmentation and downstream melanoma diagnosis. 
Using a curated dataset of 14,116 dermoscopic images from HAM10000 and MILK10K across five lesion classes, we evaluate the two representative generative paradigms: StyleGAN2-ADA and Denoising Diffusion Probabilistic Models (DDPMs) under basic geometric augmentation and advanced artifact removal pipelines. 
Synthetic melanoma images are assessed using established perceptual and distributional metrics (FID, KID, IS), feature space analysis, and their impact on diagnostic performance across five downstream classifiers. 
Experimental results demonstrate that generative architecture choice has a stronger influence on both image fidelity and diagnostic utility than preprocessing complexity. 
StyleGAN2-ADA consistently produced synthetic images more closely aligned with real data distributions, achieving the lowest FID ($\approx65.5$) and KID ($\approx0.05$), while diffusion models generated higher variance samples at the cost of reduced perceptual fidelity and class anchoring. 
Advanced artifact removal yielded only marginal improvements in generative metrics and provided limited downstream diagnostic gains, suggesting possible suppression of clinically relevant texture cues. 
In contrast, synthetic data augmentation substantially improved melanoma detection with $8$–$15\%$ absolute gains in melanoma F1-score, and ViT-B/16 achieving F1$\approx0.88$ and ROC-AUC$\approx0.98$, representing an improvement of approximately $14\%$ over non-augmented baselines. 
Our code can be found at \href{https://github.com/adarsh-crafts/SkinGenBench}{https://github.com/adarsh-crafts/SkinGenBench}
\end{abstract}

\begin{keyword}
Medical Image Synthesis  \sep Generative Adversarial Networks \sep Diffusion Models \sep Melanoma Detection \sep Synthetic Data Augmentation.
%% keywords here, in the form: keyword \sep keyword

%% PACS codes here, in the form: \PACS code \sep code

%% MSC codes here, in the form: \MSC code \sep code
%% or \MSC[2008] code \sep code (2000 is the default)

\end{keyword}

\end{frontmatter}

%%
%% Start line numbering here if you want
%%
% \linenumbers
%%%%%%%%%%%%%%%%%%%%%%%%%%%%%%%%%%%%%%%%%%%%%%%%%%%%%%%%%%%%
\section{Introduction} 
\begin{figure}[!h]
\centering
\includegraphics[scale=0.9]{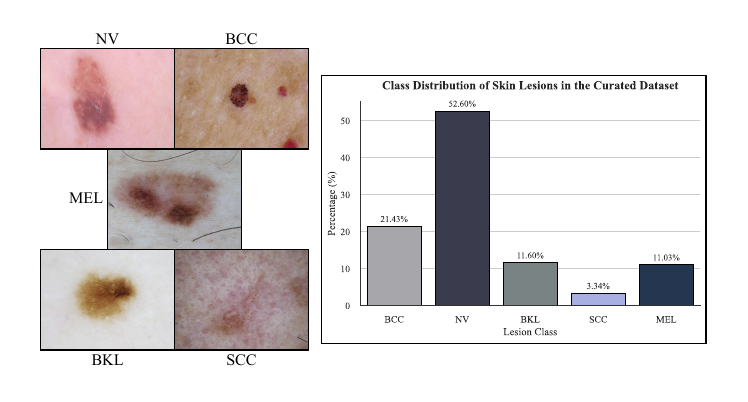}
\caption{The types of skin lesions used in this study and their distribution in the curated dataset. Namely Nevus (NV, 52.60\%), Basal Cell Carcinoma (BCC, 21.43\%), Benign Keratosis Like (BKL, 11,60\%), Melanoma (MEL, 11.03\%), and Squamous Cell Carcinoma (SCC, 3.34\%).} \label{fig: lesion-types-and-distribution}
\end{figure}

Skin cancer remains the most prevalent malignancy globally, surpassing all other cancer types combined and posing a major public health challenge. Among lighter-skinned populations, Nonmelanoma Skin Cancer (NMSC) is the most common form~\cite{lomas_systematic_2012}, while malignant melanoma (MM) is the deadliest, projected to rise by 62\% by 2040, affecting nearly half a million individuals~\cite{cassidy_analysis_2022}. The five lesion classes analyzed in this work are shown in Figure~\ref{fig: lesion-types-and-distribution}.

Early and precise detection is critical as melanoma’s five-year survival rate exceeds 99\% when identified early but drops below 32\% once metastasized~\cite{esteva_dermatologist-level_2017}. Dermoscopy enables visualization of subsurface lesion structures but remains highly dependent on clinician expertise. Deep learning, especially convolutional neural networks (CNNs), has achieved dermatologist-level classification performance~\cite{cassidy_analysis_2022,luschi_advancing_2025}. Yet, challenges persist due to severe class imbalance, limited labeled malignant samples, and data artifacts that bias models~\cite{luschi_advancing_2025,mirikharaji_survey_2023}. Public datasets like ISIC contain under 10\% malignant lesions and suffer from duplication, ruler marks, and other artifacts, limiting generalizability~\cite{cassidy_analysis_2022}.

Generative models address these limitations by learning data distributions to synthesize high-fidelity images~\cite{abeytunga_acstylegan_2025}. GAN-based methods have effectively mitigated class imbalance~\cite{pedersen_panogan_2025}, while diffusion models exhibit superior diversity and robustness in low-data regimes through iterative denoising~\cite{wang_review_2021,luschi_advancing_2025,jimenez-perez_dino-diffusion_2026}. However, their comparative performance remains underexplored, particularly regarding how preprocessing complexity (basic geometric augmentation versus aggressive artifact removal) affects generative quality and downstream diagnostic accuracy.
This work introduces SkinGenBench, the first systematic benchmark analyzing the interplay between preprocessing complexity and generative architecture choice for dermoscopic image synthesis and melanoma classification. The key contributions are:
(1) A controlled evaluation of preprocessing complexity across GANs (StyleGAN2-ADA) and diffusion models (DDPMs).
(2) A reproducible benchmarking framework comparing both paradigms under matched training conditions.
(3) A unified assessment combining generative quality metrics (FID, IS, KID) with downstream CNN and transformer-based diagnostic performance.
(4) Empirical findings demonstrating that generative architecture choice outweighs preprocessing complexity, establishing GAN-based augmentation as a robust, clinically aligned standard for melanoma detection.

\section{Literature Review}
Large public datasets such as ISIC and HAM10000 have driven progress in automated skin lesion analysis~\cite{mirikharaji_survey_2023}, yet suffer from persistent quality and representational limitations. Common artifacts (e.g., hair and ruler marks) significantly impair segmentation and classification~\cite{cassidy_analysis_2022}, and although preprocessing methods provide filter-based removal strategies~\cite{kasmi_sharprazor_2023}, these issues remain unresolved and may suppress diagnostically relevant signal. Additionally, the overrepresentation of fair-skinned populations limits model generalizability~\cite{lomas_systematic_2012}. To address data scarcity and class imbalance, generative augmentation has been widely adopted: class-conditioned GANs such as ACStyleGAN improve minority-class performance~\cite{abeytunga_acstylegan_2025}, StyleGAN2 generates melanoma images indistinguishable from real samples~\cite{luschi_advancing_2025}, and attention-based EA-GANs enhance synthesis of diagnostically salient regions, boosting classifier accuracy and F1-scores~\cite{g_class-specific_2025}. More recently, diffusion models have demonstrated superior sample quality over GANs at the cost of slower sampling~\cite{dhariwal_diffusion_2021} and self-supervised approaches such as DiNO-Diffusion further mitigate annotation constraints by conditioning latent diffusion models on DiNO embeddings, enabling high-fidelity synthesis and improved classification in low-data regimes~\cite{jimenez-perez_dino-diffusion_2026}.

\section{Methodology}
\begin{figure}[!h]
\centering
\includegraphics[scale=0.8]{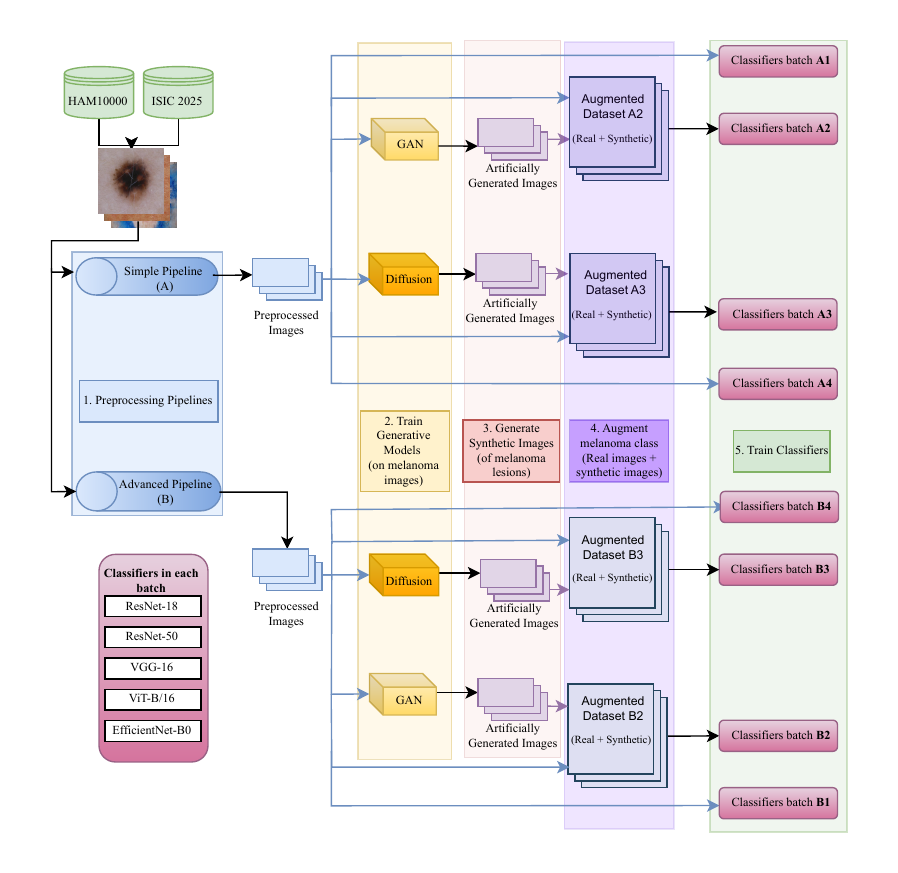}
\caption{Overall Design of the experimental study.} \label{fig:overall-flowchart}
\end{figure}

The design of this study is shown in Figure~\ref{fig:overall-flowchart}. The workflow begins with the two preprocessing pipelines (Simple pipeline A and Advanced pipeline B) through which the curated set of dermoscopic images are processed. Subsequently, generative adversarial network (GANs) and diffusion models are trained to synthesize realistic melanoma images, addressing the data imbalance. The synthetic images are augmented with the real melanoma data to construct the new augmented datasets (A2, A3, B2, B3). Pipelines A4 and B4 perform the standard augmentation used in image processing tasks without synthesized images. These datasets are then used to train and benchmark a diverse set of deep learning classifiers (ResNet18, RensNet50, VGG16, ViT-B/16, EfficientNet-B0) to evaluate the impact on diagnostic accuracy of melanoma.

\subsection{Dataset development}
This study uses an aggregated dataset of dermoscopic images of skin lesions from two well established public sources: the International Skin Imaging Collaboration (ISIC) 2025 dataset~\cite{milk_study_team_milk10k_nodate}, which contains 5,421 images, and the Human Against Machine with 10000 training images (HAM10000)~\cite{tschandl_ham10000_2018} dataset, which contains 10,015 images.
From the aggregated pool, the lesion types with insufficient sample sizes for model training were excluded, and a targeted dataset was created by including only five relevant classes. This final dataset consisted of a total of 14,116 dermoscopic images. The class-wise distribution is shown in Figure~\ref{fig: lesion-types-and-distribution}. Melanoma (MEL) is the most aggressive form of skin cancer~\cite{cassidy_analysis_2022} and is significantly underrepresented, with 1,563 images (11.03\%) in the dataset. This study focuses on improving its diagnostic accuracy.

\subsection{Preprocessing pipelines}
\begin{figure*}[!h]
\centering
\includegraphics[width=\textwidth]{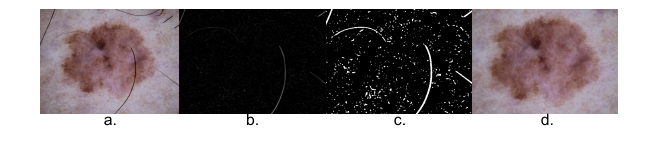}
\caption{\textbf{Stages of the Dullrazor algorithm.} Stages of the algorithm are: (a) Original image with artifacts. (b) Blackhat mask. (c) Binary mask. (d) Final image free of artifacts.} \label{fig5}
\end{figure*}

The curated dataset underwent two different preprocessing pipelines to evaluate the impact of input characteristics on the generative models. Two distinct preprocessing pipelines were created to establish separate training sets for the generative models. The first, basic pipeline employed standard built-in data augmentations such as random rotations and horizontal/vertical flipping after resizing the images to 256x256 resolution
The second, advanced pipeline integrated all the transformations from the basic pipeline but was first subjected to an artifact removal process. This additional process was designed to digitally eliminate common artifacts present in dermoscopic images, such as hair and ruler marks. The  Dullrazor algorithm was employed to achieve this crucial preprocessing step. In this technique, the dermoscopic images are converted into grayscale, then, a morphological black-hat transformation is applied to isolate dark, hair-like structures from the lighter skin background~\cite{lee_dullrazor_1997}. A binary mask is then created by thresholding the output, this mask precisely delineates the hair locations. Finally, the generated mask is used to reconstruct the pixels obscured by hair with textures approximated from the surrounding lesion and skin pixels, resulting in a clean, artifact free image. An example of the dullrazor algorithm is shown in Figure~\ref{fig5}.
The goal was to provide the generative models with a training set containing fewer irrelevant features that might deter their generative capabilities. The resulting two datasets served as the basis for training instances of the generative models, which enable a controlled assessment of how these two processing pipelines impact the quality of synthetic data augmentation in downstream classifiers

\subsection{Construction and image generation of StyleGAN2-ADA and DDPM}
GANs consist of a generator and a discriminator operating on a latent space of high-dimensional random vectors. During adversarial training, the generator transforms noise samples into synthetic images, while the discriminator differentiates them from real data, iteratively improving image realism~\cite{goodfellow_generative_2014}.In StyleGAN2, the mapping network learns disentangled latent representations, capturing independent feature variations~\cite{karras_analyzing_2020}. The StyleGAN2-ADA variant integrates Adaptive Discriminator Augmentation to counteract overfitting on limited data by dynamically applying stochastic augmentations during training~\cite{karras_training_2020}. Each input latent vector passes through eight fully connected layers to yield intermediate latent vectors that modulate multi-scale convolutional kernels. Starting from a constant $4×4×256$ tensor, the network progressively upsamples and injects style parameters to generate $256×256$ images. A dual-branch modulation controls both global and local features, reducing artifacts and enhancing diversity~\cite{zeng_expert-guided_2025}. 
Overfitting is monitored via discriminator-based metrics ${r_v}$ and ${r_t}$:

\noindent
\begin{minipage}{.5\textwidth}
  \begin{equation}
    r_v=\frac{E\left[D_{\mathrm{train}}\right]-E\left[D_{\mathrm{validation}}\right]}{E\left[D_{\mathrm{train}}\right]-E\left[D_{\mathrm{generated}}\right]}
    \label{eq:1}
  \end{equation}
\end{minipage}%
\begin{minipage}{.5\textwidth}
  \begin{equation}
    r_t=E\left[\mathrm{sign}\left(D_{\mathrm{train}}\right)\right]
    \label{eq:2}
  \end{equation}
\end{minipage}
Equation~\ref{eq:1} quantifies overfitting using training and validation sets, and Equation~\ref{eq:2} estimates overfitting from training outputs alone, adjusting augmentation probability ${p}$ every four minibatches to maintain a target threshold.
Two StyleGAN2-ADA instances were fine-tuned on separate preprocessing pipelines (Fig.~\ref{fig:overall-flowchart}), initialized from a pre-trained FFHQ network $(512\times512)$. Each was trained for $240k$ images using Adam $(lr=0.0025)$, minibatch size $16$, on an NVIDIA L4 $(22 GB)$. Augmentations included horizontal mirroring, with ADA target ${r_t}=0.7$ to prevent discriminator overfitting.

Diffusion models (DDPMs) employ a forward process that progressively adds Gaussian noise to an image and a learned reverse process that denoises it~\cite{ho_denoising_2020}. The forward process given in Equation~\ref{eq:3} converts ${x_0}$ into noisy ${x_t}$ over ${T}$ timesteps where ${\alpha}$ denotes the cumulative noise coefficient:
\begin{equation}
x_t = \sqrt{\alpha_t}\,x_0 + \sqrt{1 - \alpha_t}\,\epsilon_t, \quad 
\epsilon_t \sim \mathcal{N}(0, I)
\label{eq:3}
\end{equation}

The reverse process reconstructs ${x_0}$ using a neural network ${\epsilon_\theta(x_t,t)}$ within a Guassian transition as shown in Equation~\ref{eq:4} where ${\mu(x_t,t)}$ is the predicted mean and $(\sigma_t^2)$ is the variance that controls the level of uncertainty at each timestep \( t \)~\cite{wang_diffusion_2025}:
\begin{equation}
p_\theta\left(x_{t-1}\mid x_t\right)=\mathcal{N}\left(x_{t-1};\mu_\theta\left(x_t,t\right),\sigma_t^2I\right)
\label{eq:4}
\end{equation}
The model adopts a U-Net backbone with self-attention and residual connections~\cite{vaswani_attention_2023}, minimizing MSE between predicted and true noise~\cite{jain_panodiff-sr_2025}.
Two DDPMs were fine-tuned from a CELEBA-HQ pretrained checkpoint $(256×256)$ for $120$ epochs (matching the GAN’s effective image passes). Training used Adam $(lr=1\times10^{-5})$, $500$ warm-up steps, EMA weight averaging, and FP16 precision with minibatch size $16$. Random horizontal flipping and center cropping ensured consistent feature learning~\cite{g_class-specific_2025}.

\subsection{Image classification using downstream classifiers}
To get a performance perspective regarding the impact of augmentation of synthetic data generated from the generative models, we trained five classifiers, namely, ResNet-18, ResNet-50, VGG-16, Vit-B/16, and EfficientNet-B0 to distinguish between the types of skin lesions. Transfer learning was employed for all the classification models with standard ImageNet-1K pretrained weights. During training, the models were finetuned on images from the augmented datasets using the same optimizer (Stochastic Gradient Descent) with a learning rate of $0.001$, momentum of $0.9$, and weight decay of 0.1. Training was conducted for $30$ epochs using a Cross-Entropy loss function. A MultiStepLR scheduler was applied with milestones at epochs $15$ and $25$ and a decay factor of $0.1$. Data augmentations during training cropping images to $224\times224$ resolution, random horizontal flips, and normalization.

\subsection{Systematic Evaluation of Generative Models Across Preprocessing Pipelines}
We begin by detailing the training protocol and architectural settings used for the GAN model.
Both training runs across the two preprocessing pipelines were conducted without transfer learning and instead trained the GAN architecture from scratch using the augmented melanoma dataset. Images were resized to $256\times256$ pixels and normalized to the range $[-1,1]$. Training was performed using the Adam optimizer with a learning rate of $0.0002$ and momentum parameters $(\beta_{1}=0.5, \beta_{2}=0.999)$. The model was trained for a total of $50,000$ update steps (iterations) with a minibatch size of $32$.
Several regularization strategies were employed to stabilize the adversarial training. DiffAugment was used with the color and translation augmentation policies applied consistently to both real and synthesized images. Additionally, perceptual reconstruction losses were incorporated using LPIPS with a VGG-based “net-lin” backbone to guide the discriminator’s region-based reconstruction pathways. The generator parameters were also updated with an exponential moving average (EMA) using a decay factor of $0.999$ to improve sample quality during inference.
To ensure reproducibility, all the experiments were executed using PyTorch with random seeds. 

The diffusion model on the other hand, was trained from scratch using a DDPM with a U-Net backbone using an image resolution of $224\times224$. The network was trained for $1000$ epochs using the AdamW optimizer with a learning rate of $1\times10^{-4}$,  $(\beta_{1}=0.5, \beta_{2}=0.999)$, no weight decay, and a cosine learning-rate schedule with $300$ warmup steps. A total of $1000$ diffusion timesteps were used with a cosine beta schedule for the forward noising process. The training used FP16 mixed precision, minibatch of size $2$, and gradient accumulation over $1000$ steps.
To improve sample quality and stabilize training, we maintained a cosine-scheduled Exponential Moving Average (EMA) of the model parameters, initialized at $\gamma=0.996$ and gradually increased toward $1.0$. The EMA model was used for sample generation during training. All images were resized to $224\times224$, converted to RGB when necessary, and normalized to $[-1,1]$ range. No stochastic data augmentation was applied. At evaluation time, samples were generated by intializing from Gaussian noise and applying the learned denoising process, and the outputs were rescaled to $[0,1]$ for visualization.

\section{Results and Discussion}

We defined two sets of images for comparison: Set A underwent basic preprocessing, while Set B underwent the advanced preprocessing pipeline. For systematic comparison, as shown in Table~\ref{tab:imagesets}, we define eight subsets of images, namely, $(1)$ BSGT for Ground-Truth (i.e., real images) pre-processed using the basic pipeline, $(2)$ BSGN for melanoma images generated with GAN trained using images from the basic pipeline, $(3)$ BSDF for melanoma images generated with the diffusion model trained using images from the basic pipeline, $(4)$ ADGT for Ground-Truth (i.e., real images) pre-processed using the advanced pipeline, $(5)$ ADGN for melanoma images generated with GAN trained using images from the advanced pipeline, $(6)$ ADDF for melanoma images generated with the diffusion model trained using images from the advanced pipeline~\cite{jain_panodiff-sr_2025}, $(7)$ BSGTA for melanoma images augmented using standard image augmentation techniques (i.e. illumination normalization (Shades-of-Gray), Contrast Limited Adaptive Histogram Equalization (CLAHE), horizontal flips, rotations, random brightness adjustments, blurring, and course dropouts) using images from the basic pipeline, $(8)$ ADGTA for melanoma images augmented with the same techniques as BSGTA using images from the advanced pipeline.

\begin{table*}[!h]
\centering
\caption{Image subsets derived from the two preprocessing pipelines and generative models at epoch no. 1000.}
\label{tab:imagesets}
\begin{tabular}{lcc}
\hline
\textbf{Source} & \textbf{Basic (BS)} & \textbf{Advanced (AD)} \\
\hline
Ground Truth (GT) & BSGT & ADGT \\
StyleGAN2-ADA (GN) & BSGN & ADGN \\
DDPM (DF) & BSDF & ADDF \\
Ground Truth Augmented (GTA) & BSGTA & ADGTA \\
\hline
\end{tabular}
\end{table*}

\begin{figure}[!h]
\centering
% ---- (a) Wide image ----
\begin{subfigure}{\textwidth}
    \centering
    \includegraphics[width=\textwidth]{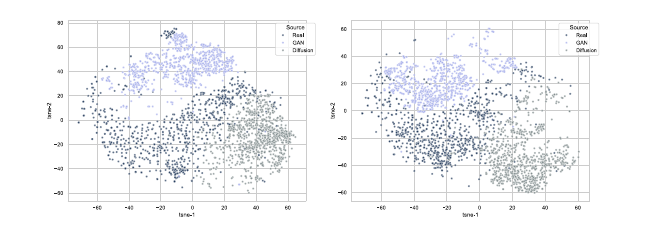}
    \caption{
    t-SNE embeddings for the basic (left) and advanced (right) pipelines,
    showing the distributions of GT, GN, and DF samples.
    }
    \label{fig:tsne-basic-advanced}
\end{subfigure}

\vspace{0.5em}

% ---- (b) Table ----
\begin{subfigure}{\textwidth}
    \centering
    \scriptsize
    \begin{tabular}{l c c c}
        \toprule
        \textbf{Pipeline} &
        \textbf{GT--GN} &
        \textbf{GT--DF} &
        \textbf{GN--DF} \\
        \midrule
        Advanced & 34.4711 & 51.4366 & 74.3892 \\
        Basic    & 48.5551 & 51.0257 & 75.6382 \\
        \bottomrule
    \end{tabular}
    \caption{
    Euclidean distances between cluster centers in t-SNE space.
    }
    \label{tab:euclidean-distance}
\end{subfigure}
\caption{Qualitative and quantitative analysis of feature-space distributions across preprocessing pipelines.}
\label{fig:tsne-and-distances}
\end{figure}

The t-SNE plot of features extracted using ResNet-50 was created to visualize the distribution of real and synthetic melanoma images pre-processed by the two pipelines, as shown in Figure~\ref{fig:tsne-basic-advanced}. This provides an informative visual of how real, GAN-generated, and Diffusion generated images are distributed in a reduced feature space. The embeddings reveal distinct clusters rather than a homogeneous mixture of all sources in either preprocessing condition. The BSGN samples form a well separated, compact and coherent cluster whereas BSDF samples are more broadly spread. This suggests that GAN's latent space has learned a relatively constrained representation under Basic preprocessing, which aligns with prior findings in \cite{frid-adar_gan-based_2018}. This structure is quantitatively supported by the Euclidean distances between the distributions shown in Table~\ref{tab:euclidean-distance}, where the distance between GT--GN $(48.56)$ is slightly smaller than GT--DF $(51.03)$. In contrast, ADDF samples show greater dispersion across the embedding space, having the largest separation from GANs $(75.64)$, reflecting diffusion models’ tendency to generate higher-variance, multimodal outputs. The GT--GN samples have a substantially smaller distance $(34.47)$ compared to GT--DF $(52.44)$. These findings align with recent studies showing that diffusion models can overfit on small medical datasets and may generalize less effectively GANs in certain augmentation regimes\cite{akbar_beware_2024} while systematic reviews continue to highlight the practicality, stability, and strong performance GAN-based augmentation in medical imaging\cite{hussain_generative_2025}. Comparative evaluations further show that GANs often offer more controllable variability than diffusion models, which may explain why data augmentation with the structured, low noise manifold of BSGN samples yield the strongest classifier performance in our experiments as shown in Table~\ref{tab: condensed_global_metrics}.

\begin{figure}[!h]
\centering
% ---- Row 1 ----
\begin{subfigure}[t]{0.45\textwidth}
    \centering
    \includegraphics[width=\linewidth]{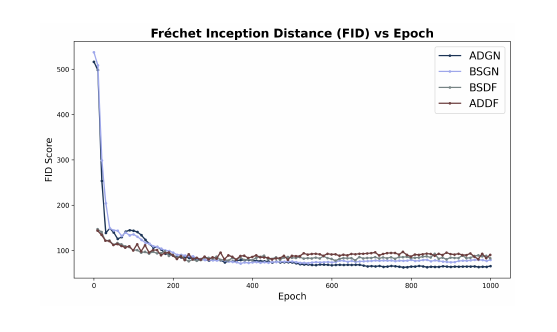}
    \caption{Figure 1}
\end{subfigure}
\hfill
\begin{subfigure}[t]{0.45\textwidth}
    \centering
    \includegraphics[width=\linewidth]{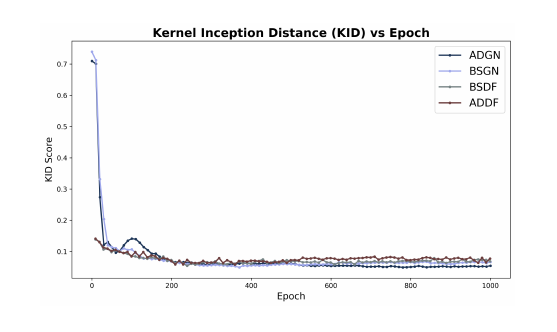}
    \caption{Figure 2}
\end{subfigure}

\vspace{0.5cm}

% ---- Row 2 ----
\begin{subfigure}[t]{0.45\textwidth}
    \centering
    \includegraphics[width=\linewidth]{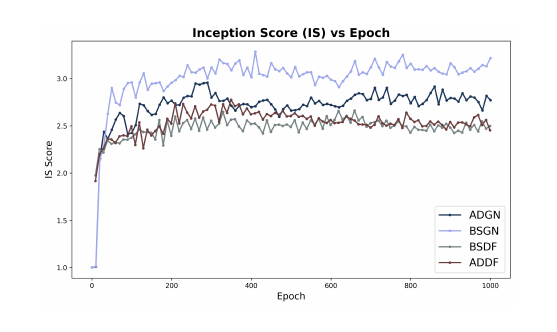}
    \caption{Figure 3}
\end{subfigure}
\hfill
\begin{subfigure}[t]{0.45\textwidth}
    \centering
    \raisebox{1.2cm}{ % <-- adjust this value
        \begin{minipage}{\linewidth}
        \small
        \centering
        \begin{tabular}{lccc}
            \toprule
            \textbf{Model} & \textbf{FID} & \textbf{KID} & \textbf{IS} \\
            \midrule
            BSGN & 79.3582 & 0.0664 & 3.2165 \\
            ADGN & \textbf{65.4739} & \textbf{0.0546} & 2.7726 \\
            BSDF & 83.0421 & 0.0684 & 2.4995 \\
            ADDF & 90.2158 & 0.0772 & 2.4542 \\
            \bottomrule
            \label{tab: fid-kid-is}
        \end{tabular}
        \caption{Quantitative evaluation of synthesized samples at epoch 1000 across preprocessing pipelines}
        \end{minipage}
    }

\end{subfigure}
\caption{Image Sample Quality Metrics}
\label{fig:img-quality-combined}
\end{figure}

Furthermore, we evaluated the generated melanoma images against their corresponding lesion images in the original dataset using Fréchet Inception Distance (FID), Inception Score (IS), and Kernel Inception Distance (KID), as shown in Figure~\ref{fig:img-quality-combined}. FID measures the similarity between the real and generated image distributions by comparing their feature vector embeddings from a pretrained Inception network, with lower values indicating higher quality and diversity~\cite{heusel_gans_2018}. IS evaluates the realism and diversity of the generated images by measuring the prediction confidence and distribution variance using a pretrained Inception model, where higher scores reflect more realistic and varied images~\cite{salimans_improved_2016}.
From the table in Figure~\ref{fig:img-quality-combined}, the Fréchet Inception Distance (FID) scores for diffusion-generated images ($BSDF=83.04$, $ADDF=90.22$) were higher than those of GAN-generated images ($BSGN=79.36$, $ADGN=65.47$), indicating that GANs produced images closer to the real data distribution. The notably lower FID values of GAN-generated images suggest better fidelity and closer alignment with real melanoma lesion features. Moreover, a minor improvement in FID was observed when advanced preprocessing was used, confirming that artifact removal and preprocessing refinement marginally improved image realism. Diffusion models generated more diverse samples and yet exhibited higher FID values due to broader sampling of the latent space, producing overly smoothed textures compared to GANs. The Inception Scores reported in the Figure~\ref{fig:img-quality-combined}, show that the GAN generated images $(BSGN=3.22, ADGN=2.77)$, and the diffusion model generated images yielded lower values $(BSDF=2.50, ADDF=2.45)$, reflecting reduced class confidence and diversity compared to the image samples generated by GAN. In summary, GANs outperformed diffusion models across FID, IS, and KID metrics, producing samples that we more realistic, more class coherent, and more closely matched to the underlying melanoma distribution, while diffusion models at the expense of sharpness and fine lesion details, provided more structural diversity.

\subsection{Classification of the skin lesion images}
From Table~\ref{tab: condensed_global_metrics}, the classifier results demonstrated consistent trends favoring Pipeline A over pipeline B across all configurations, independent of the classifier architecture and the generative settings. The A2 pipeline achieved consistently strong global performance with top models reaching macro-F1 scores in the range of $\approx 0.80$-$0.84$ and overall accuracy approaching $\approx0.90$, indicating stable gains across architectures rather than isolated improvements.
\begin{table*}[!h]
\centering
\scriptsize
\setlength{\tabcolsep}{3pt}
\renewcommand{\arraystretch}{0.9}
\caption{Global classification metrics across models and preprocessing pipelines.}
\begin{tabular}{llcccccccc}
\toprule
\textbf{Model} & \textbf{Metric} & A1 & A2 & A3 & A4 & B1 & B2 & B3 & B4 \\
\midrule

\multirow{6}{*}{efficientnet\_b0}
& F1      & 0.7504 & 0.7977 & 0.7941 & 0.7452 & 0.7251 & 0.7878 & 0.7771 & 0.7256 \\
& B.Acc   & 0.7405 & 0.7871 & 0.7825 & 0.7267 & 0.7141 & 0.7795 & 0.7697 & 0.7134 \\
& MCC     & 0.7621 & 0.8033 & 0.7937 & 0.7622 & 0.7433 & 0.7878 & 0.7846 & 0.7423 \\
& ROC-AUC & 0.9604 & 0.9698 & 0.9700 & 0.9618 & 0.9568 & 0.9665 & 0.9657 & 0.9562 \\
& Acc     & 0.8458 & 0.8657 & 0.8590 & 0.8472 & 0.8349 & 0.8551 & 0.8532 & 0.8338 \\
& Brier   & 0.0446 & 0.0402 & 0.0407 & 0.0440 & 0.0475 & 0.0434 & 0.0436 & 0.0476 \\

\midrule
\multirow{6}{*}{resnet18}
& F1      & 0.6996 & 0.7525 & 0.7530 & 0.7041 & 0.6897 & 0.7386 & 0.7426 & 0.6950 \\
& B.Acc   & 0.6754 & 0.7424 & 0.7391 & 0.6817 & 0.6682 & 0.7201 & 0.7284 & 0.6716 \\
& MCC     & 0.7311 & 0.7594 & 0.7658 & 0.7274 & 0.7141 & 0.7546 & 0.7539 & 0.7186 \\
& ROC-AUC & 0.9541 & 0.9611 & 0.9641 & 0.9521 & 0.9455 & 0.9565 & 0.9577 & 0.9452 \\
& Acc     & 0.8278 & 0.8360 & 0.8405 & 0.8250 & 0.8169 & 0.8331 & 0.8325 & 0.8197 \\
& Brier   & 0.0497 & 0.0479 & 0.0459 & 0.0499 & 0.0532 & 0.0495 & 0.0482 & 0.0529 \\

\midrule
\multirow{6}{*}{resnet50}
& F1      & 0.8001 & \textbf{$0.8393$} & 0.8371 & 0.8020 & 0.7909 & 0.8334 & 0.8291 & 0.8003 \\
& B.Acc   & 0.7919 & \textbf{$0.8433$} & 0.8334 & 0.7898 & 0.7876 & 0.8221 & 0.8212 & 0.7976 \\
& MCC     & 0.8141 & 0.8525 & \textbf{$0.8526$} & 0.8221 & 0.8072 & 0.8433 & 0.8469 & 0.8120 \\
& ROC-AUC & 0.9756 & 0.9802 & 0.9815 & 0.9762 & 0.9724 & 0.9798 & 0.9799 & 0.9725 \\
& Acc     & 0.8797 & \textbf{$0.8989$} & \textbf{$0.8992$} & 0.8850 & 0.8740 & 0.8931 & 0.8953 & 0.8776 \\
& Brier   & 0.0355 & 0.0314 & 0.0314 & 0.0354 & 0.0379 & 0.0324 & 0.0317 & 0.0363 \\

\midrule
\multirow{6}{*}{vgg16}
& F1      & 0.7869 & 0.8181 & 0.8061 & 0.7932 & 0.7627 & 0.8009 & 0.7993 & 0.7657 \\
& B.Acc   & 0.7834 & 0.8167 & 0.8051 & 0.7938 & 0.7615 & 0.8011 & 0.8014 & 0.7589 \\
& MCC     & 0.7931 & 0.8243 & 0.8112 & 0.7953 & 0.7734 & 0.8021 & 0.8085 & 0.7795 \\
& ROC-AUC & 0.9698 & 0.9774 & 0.9767 & 0.9683 & 0.9633 & 0.9734 & 0.9739 & 0.9643 \\
& Acc     & 0.8659 & 0.8797 & 0.8708 & 0.8677 & 0.8525 & 0.8644 & 0.8689 & 0.8571 \\
& Brier   & 0.0394 & 0.0365 & 0.0367 & 0.0398 & 0.0439 & 0.0396 & 0.0385 & 0.0433 \\

\midrule
\multirow{6}{*}{vit\_b\_16}
& F1      & 0.8096 & \textbf{0.8393} & 0.8266 & 0.7997 & 0.7932 & 0.8268 & 0.8378 & 0.7998 \\
& B.Acc   & 0.7931 & 0.8348 & 0.8198 & 0.7885 & 0.7773 & 0.8134 & 0.8223 & 0.7852 \\
& MCC     & 0.8230 & 0.8515 & 0.8434 & 0.8160 & 0.8093 & 0.8398 & 0.8421 & 0.8161 \\
& ROC-AUC & 0.9753 & \textbf{0.9822} & 0.9814 & 0.9750 & 0.9746 & 0.9807 & 0.9815 & 0.9723 \\
& Acc     & 0.8857 & 0.8985 & 0.8928 & 0.8807 & 0.8765 & 0.8906 & 0.8922 & 0.8807 \\
& Brier   & 0.0361 & \textbf{0.0302} & 0.0320 & 0.0358 & 0.0370 & 0.0332 & 0.0331 & 0.0374 \\

\bottomrule
\end{tabular}
\label{tab: condensed_global_metrics}
\end{table*}
As shown in Table~\ref{tab: condensed_mel_classification_metrics}, the augmentation of synthetic data generated by both the generative models in configurations A2 (BSGN) and A3 (BSDF) improved the mean ROC-AUC and per-class discrimination for melanoma (MEL), which is clinically critical and underrepresented. Particularly, the per-class F1 score for MEL increased substantially across architectures, with absolute gains of approximately $+0.07$–$0.12$, while also enhancing the other minority classes such as BKL and SCC, increasing the macro-F1 by up to $\approx +0.04$–$0.05$, indicating broader class balance improvements without overfitting (AUC $>0.96$ across all cases).

\begin{table*}[!h]
\centering
\scriptsize
\setlength{\tabcolsep}{1pt}
\renewcommand{\arraystretch}{0.9}
\caption{Melanoma classification performance across models and preprocessing pipelines.}
\begin{tabular}{llcccccccc}
\toprule
\textbf{Model} & \textbf{Metric} & A1 & A2 & A3 & A4 & B1 & B2 & B3 & B4 \\
\midrule

% ================= EfficientNet-B0 =================
\multirow{8}{*}{efficientnet\_b0}
& Sens    & 0.5623 & 0.7781 & 0.7602 & 0.5751 & 0.4920 & 0.7537 & 0.7357 & 0.5080 \\
& Spec    & 0.9742 & 0.9667 & 0.9635 & 0.9722 & 0.9734 & 0.9639 & 0.9695 & 0.9722 \\
& Prec    & 0.7303 & 0.8503 & 0.8351 & 0.7200 & 0.6968 & 0.8354 & 0.8542 & 0.6943 \\
& F1      & 0.6354 & 0.8126 & 0.7959 & 0.6394 & 0.5768 & 0.7925 & 0.7905 & 0.5867 \\
& F2      & 0.5894 & 0.7916 & 0.7741 & 0.5992 & 0.5227 & 0.7687 & 0.7567 & 0.5368 \\
& ROC-AUC & 0.9353 & 0.9633 & 0.9637 & 0.9392 & 0.9301 & 0.9575 & 0.9575 & 0.9313 \\
& PR-AUC  & 0.7296 & 0.9043 & 0.9048 & 0.7379 & 0.6866 & 0.8881 & 0.8898 & 0.6854 \\
& DOR     & 48.5408 & 101.7550 & 83.6967 & 47.3878 & 35.4751 & 81.7015 & 88.3633 & 36.1511 \\

\midrule
% ================= ResNet-18 =================
\multirow{8}{*}{resnet18}
& Sens    & 0.5048 & 0.7390 & 0.7423 & 0.5176 & 0.4952 & 0.7325 & 0.7096 & 0.5208 \\
& Spec    & 0.9730 & 0.9576 & 0.9659 & 0.9667 & 0.9667 & 0.9627 & 0.9667 & 0.9710 \\
& Prec    & 0.6991 & 0.8089 & 0.8410 & 0.6585 & 0.6485 & 0.8269 & 0.8382 & 0.6907 \\
& F1      & 0.5863 & 0.7724 & 0.7886 & 0.5796 & 0.5616 & 0.7768 & 0.7686 & 0.5938 \\
& F2      & 0.5345 & 0.7520 & 0.7601 & 0.5407 & 0.5198 & 0.7496 & 0.7321 & 0.5477 \\
& ROC-AUC & 0.9243 & 0.9542 & 0.9557 & 0.9229 & 0.9170 & 0.9496 & 0.9505 & 0.9157 \\
& PR-AUC  & 0.6696 & 0.8774 & 0.8894 & 0.6719 & 0.6404 & 0.8689 & 0.8796 & 0.6386 \\
& DOR     & 36.7717 & 63.8751 & 81.5370 & 31.1254 & 28.4610 & 70.6878 & 70.8999 & 36.4405 \\

\midrule
% ================= ResNet-50 =================
\multirow{8}{*}{resnet50}
& Sens    & 0.7444 & 0.8401 & 0.8548 & 0.6869 & 0.6933 & 0.8320 & 0.8418 & 0.6965 \\
& Spec    & 0.9655 & 0.9758 & 0.9734 & 0.9782 & 0.9655 & 0.9718 & 0.9758 & 0.9679 \\
& Prec    & 0.7281 & 0.8941 & 0.8866 & 0.7963 & 0.7138 & 0.8778 & 0.8943 & 0.7291 \\
& F1      & 0.7362 & 0.8663 & 0.8704 & 0.7376 & 0.7034 & 0.8543 & 0.8672 & 0.7124 \\
& F2      & 0.7411 & 0.8504 & 0.8610 & 0.7063 & 0.6973 & 0.8408 & 0.8518 & 0.7028 \\
& ROC-AUC & 0.9642 & 0.9787 & 0.9790 & 0.9649 & 0.9598 & 0.9778 & 0.9802 & 0.9596 \\
& PR-AUC  & 0.8314 & 0.9445 & 0.9446 & 0.8284 & 0.8018 & 0.9410 & 0.9455 & 0.8110 \\
& DOR     & 81.4830 & 211.9271 & 215.6458 & 98.3655 & 63.2397 & 170.8601 & 214.5276 & 69.1254 \\

\midrule
% ================= VGG-16 =================
\multirow{8}{*}{vgg16}
& Sens    & 0.6326 & 0.8108 & 0.8026 & 0.6262 & 0.6102 & 0.7651 & 0.7896 & 0.6070 \\
& Spec    & 0.9695 & 0.9730 & 0.9691 & 0.9710 & 0.9643 & 0.9675 & 0.9702 & 0.9710 \\
& Prec    & 0.7200 & 0.8796 & 0.8632 & 0.7286 & 0.6797 & 0.8512 & 0.8658 & 0.7224 \\
& F1      & 0.6735 & 0.8438 & 0.8318 & 0.6735 & 0.6431 & 0.8058 & 0.8259 & 0.6597 \\
& F2      & 0.6483 & 0.8237 & 0.8140 & 0.6443 & 0.6230 & 0.7809 & 0.8037 & 0.6271 \\
& ROC-AUC & 0.9527 & 0.9729 & 0.9730 & 0.9505 & 0.9378 & 0.9653 & 0.9663 & 0.9414 \\
& PR-AUC  & 0.7724 & 0.9228 & 0.9278 & 0.7726 & 0.7175 & 0.9068 & 0.9129 & 0.7333 \\
& DOR     & 54.6484 & 154.5564 & 127.3528 & 56.1770 & 42.2879 & 96.8742 & 122.3632 & 51.8009 \\

\midrule
% ================= ViT-B/16 =================
\multirow{8}{*}{vit\_b\_16}
& Sens    & 0.6869 & \textbf{0.8564} & \textbf{0.8564} & 0.7125 & 0.6709 & 0.8222 & 0.8352 & 0.7061 \\
& Spec    & 0.9790 & \textbf{0.9798} & 0.9714 & 0.9778 & 0.9782 & 0.9786 & 0.9790 & 0.9770 \\
& Prec    & 0.8022 & \textbf{0.9115} & 0.8794 & 0.7993 & 0.7925 & 0.9032 & 0.9062 & 0.7921 \\
& F1      & 0.7401 & \textbf{0.8831} & 0.8678 & 0.7534 & 0.7266 & 0.8608 & 0.8693 & 0.7466 \\
& F2      & 0.7072 & \textbf{0.8669} & 0.8609 & 0.7283 & 0.6922 & 0.8372 & 0.8485 & 0.7218 \\
& ROC-AUC & 0.9632 & 0.9802 & 0.9811 & 0.9650 & 0.9655 & 0.9787 & \textbf{0.9820} & 0.9604 \\
& PR-AUC  & 0.8317 & \textbf{0.9511} & 0.9473 & 0.8415 & 0.8216 & 0.9405 & 0.9483 & 0.8177 \\
& DOR     & 102.1602 & \textbf{288.9372} & 202.9238 & 109.0665 & 91.4139 & 211.2416 & 236.0575 & 102.0096 \\
\bottomrule
\end{tabular}
\label{tab: condensed_mel_classification_metrics}
\end{table*}

Overall, generative augmentation enhanced MEL F1-score by $8$–$15\%$ across architectures, confirming the success in improving discrimination for melanoma cases. These results highlight the robustness of both convolutional and transformer-based networks as ViT-B16, remains the best performing architecture in terms of melanoma specific metrics (MEL F1$\approx0.88$, ROC-AUC$\approx0.98$), followed closely by ResNet50 , which achieved a comparable global performance with strong sensitivity and diagnostic odds ratios. 
These gains were further supported by improvements in calibration and diagnostic reliability, as reflected by lower Brier scores, higher MCC values, and consistently elevated PR-AUC and DOR for melanoma.

\subsection{Visualization of the classifiers' saliency maps}
\begin{figure}[!h]
\centering
\includegraphics[scale=0.9]{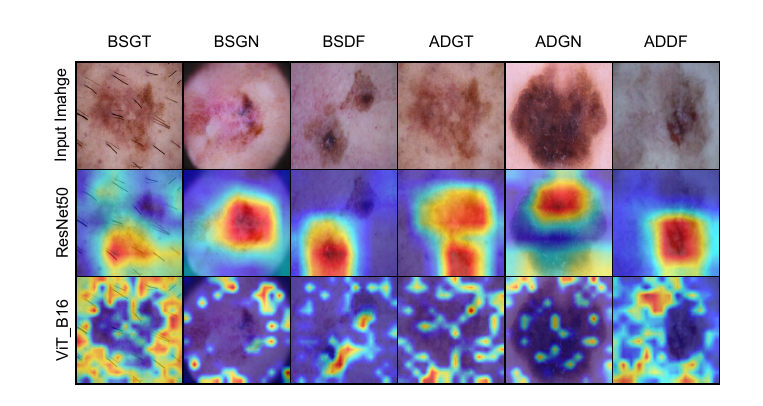}
\caption{GradCAM Visualization of the saliency maps produced by ResNet50 \& ViT-B-16 classifier.}
\label{gradcam-heatmap}
\end{figure}

As seen in Figure~\ref{gradcam-heatmap}, the Grad-CAM visualization of the saliency maps generated by ViT-B16 and ResNet50 classifiers trained on the BSGN dataset (A2 pipeline achieved the best performance) reveals interpretability trends consistent with recent dermatological AI literature. While ResNet-50 produced compact and diagnostically coherent saliency regions aligned with the pigmented lesion boundary, ViT-B16 demonstrated a patch-level diffusion of attention, with dispersed activations beyond the lesion perimeter. This is consistent with recent evidence that transformer models, being effective in capturing global context, tend to exhibit lower spatial precision in medical imaging~\cite{patricio_explainable_2023,ertmer_domain_2023}. BSGN, BSDF, ADGT, ADGN, and ADDF samples reflected greater irregular and spatially fragmented heatmaps, especially for ViT-B16. This can be attributed to high-frequency texture artifacts inherent to artifact removal algorithms, GAN \& Diffusion synthesis introducing adversarial noise that disrupts transformer attention stability as previously identified in~\cite{luschi_advancing_2025} and~\cite{abeytunga_acstylegan_2025}. The ADDF samples showed smoother and more anatomically plausible activations closer to BSGT activations, particularly in ViT-B16, supporting recent findings that diffusion-based augmentation yields structurally coherent and interpretable activations~\cite{jain_panodiff-sr_2025,wang_diffusion_2025}.

\section{Conclusion and Future Works}
This study introduced \textit{SkinGenBench}, a reproducible benchmark for evaluating how preprocessing complexity interacts with generative model choice in synthetic dermoscopic image generation and melanoma classification. Through controlled experiments across preprocessing pipelines, generative architectures (StyleGAN2-ADA, DDPM), and five representative CNN and transformer classifiers, we derived several clinically relevant insights. First, generative architecture had a stronger impact on image fidelity and diagnostic accuracy than preprocessing complexity. StyleGAN2-ADA achieved the most realistic outputs (FID$\approx65.5$ under ADGN) and lowest KID, while DDPMs offered higher diversity but weaker class anchoring and perceptual fidelity, confirmed by t-SNE dispersion. Second, advanced preprocessing (e.g., DullRazor) yielded only marginal gains (slightly lowering FID/KID yet sometimes removing diagnostically useful texture) indicating it should be applied conservatively. Model architecture and augmentation strategy dominated performance effects. Third, synthetic augmentation notably enhanced melanoma detection: MEL F1 improved by $8$–$15$\%, with consistent increases in ROC-AUC$>0.96$, PR-AUC, and diagnostic odds ratios. Transformer classifiers, especially ViT-B/16, delivered the best melanoma-specific metrics ($F1\approx0.88$, $ROC-AUC\approx0.98$), while ResNet-50 maintained strong calibration (low Brier score) and sensitivity, underscoring the reliability of GAN-based augmentation for class imbalance mitigation.
Interpretability analyses revealed trade-offs: ResNet-50 produced compact, morphologically aligned Grad-CAM saliency maps ideal for trust calibration, while ViT-B/16 showed broader, context-sensitive patterns. Diffusion-augmented data yielded smoother and anatomically coherent attention maps, suggesting benefits for hybrid interpretability auditing.

Future work will extend SkinGenBench toward:
$(1)$ exploring conditional and self-supervised diffusion models (e.g., DiNO-Diffusion, Stable Diffusion variants) for controllable generation.
$(2)$ incorporating human-in-the-loop dermatological evaluation to complement automated realism metrics.
$(3)$ developing quantitative interpretability scores tied to lesion regions for clinical faithfulnes, and
$(4)$ improving fairness and generalization via inclusion of darker skin tones and multi-institutional datasets.
Overall, SkinGenBench provides practical, reproducible guidance for synthetic dermatologic image generation and emphasizes that clinically aligned augmentation requires a balanced pursuit of realism, diversity, interpretability, and diagnostic value, rather than maximizing any single metric.

\section*{Acknowledgment}
I would like to express my sincere gratitude to Sanyam Jain and Prof. Jeba Shiney O for their continuous guidance, valuable feedback, and support throughout this project. I also extend my special thanks to Sanyam Jain for serving as the corresponding author and for providing essential resources for experimentation and publication, including access to GPU resources on Amazon EC2 (g4dn.2xlarge and g4dn.4xlarge instances). This work was carried out as part of my master’s degree at Alliance University and did not receive any external funding.
%%%%%%%%%%%%%%%%%%%%%%%%%%%%%%%%%%%%%%%%%%%%%%%%%%%%%%%%%%%%%
%% main text
% \section{}
% \label{}

%% The Appendices part is started with the command \appendix;
%% appendix sections are then done as normal sections
%% \appendix

%% \section{}
%% \label{}

%% References
%%
%% Following citation commands can be used in the body text:
%% Usage of \cite is as follows:
%%   \cite{key}         ==>>  [#]
%%   \cite[chap. 2]{key} ==>> [#, chap. 2]
%%

%% References with BibTeX database:

\bibliographystyle{elsarticle-num}
\bibliography{references}

%% Authors are advised to use a BibTeX database file for their reference list.
%% The provided style file elsarticle-num.bst formats references in the required Procedia style

%% For references without a BibTeX database:

% \begin{thebibliography}{00}

%% \bibitem must have the following form:
%%   \bibitem{key}...
%%

% \bibitem{}

% \end{thebibliography}

\end{document}